\documentclass[9pt,twocolumn,twoside]{osajnl}
\journal{ol} 
\setboolean{shortarticle}{true}

\title{Simultaneous two-photon resonant optical laser locking (STROLLing) in the hyperfine Paschen--Back regime}

\author[$\diamond$*]{Renju~S.~Mathew}
\author[$\diamond$]{Francisco~Ponciano-Ojeda}
\author[$\diamond$]{James Keaveney}
\author[$\diamond$]{Daniel~J.~Whiting}
\author[$\diamond$]{Ifan~G.~Hughes}

\affil[$\diamond$]{Joint Quantum Centre (JQC) Durham-Newcastle, Durham University, Department of Physics, South Road, Durham, DH1 3LE, United Kingdom}
\affil[*]{Corresponding author:  r.s.mathew@durham.ac.uk}

\doi{\url{https://doi.org/10.1364/OL.43.004204}}

\begin{abstract}
We demonstrate a technique to lock simultaneously two laser frequencies to each step of a two-photon transition in the presence of a magnetic field sufficiently large to gain access to the hyperfine Paschen-Back regime. 
A ladder configuration with the 5S$_{1/2}$, 5P$_{3/2}$ and 5D$_{5/2}$ terms in a thermal vapour of $^{87}$Rb atoms is used. The two lasers remain locked for more than 24 hours.
For the sum of the laser frequencies, which represents the stability of the two-photon lock, we measure a frequency instability of less than the Rb D$_2$ natural linewidth of 6 MHz for nearly all measured time scales. 
\end{abstract}

\setboolean{displaycopyright}{true}

\begin{document}

\maketitle

\begin{figure}[t!]
	\centering
	\includegraphics[width=\linewidth]{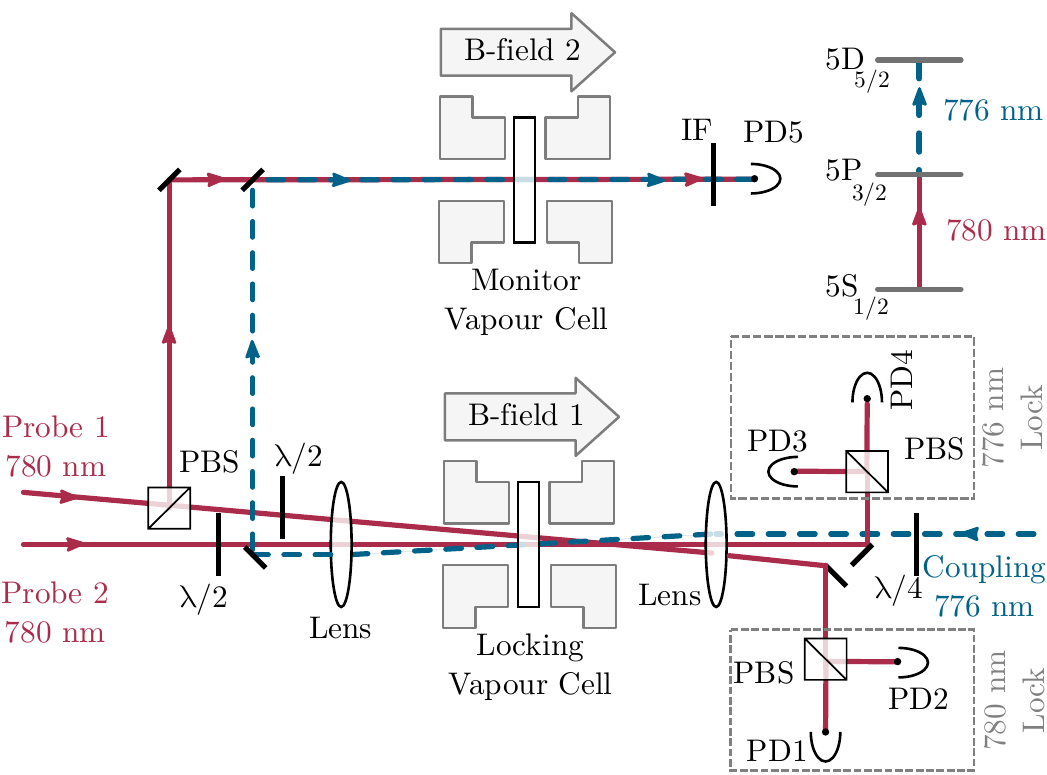}
	\caption{Schematic of the experimental configuration. The probe beams (red solid lines) and coupling beam (blue dashed line) are counterpropagated and focussed through the locking vapour cell with a path length of 1 mm containing isotopically enriched $^{87}$Rb in a uniform magnetic field of up to 0.6~T along the beam axis. Only probe beam 2 and the coupling beam are overlapped within the cell. Angles are not to scale. The beam polarisations are set by a half- and quarter- waveplates ($\lambda/2$ \& $\lambda/4$). The Stokes parameter $S_1$ is measured by subtracting the signals from the photodiodes (PD) at the output of a polarising beam splitter (PBS). PDs 1 \& 2 are for the 780 nm lock and 3 \& 4 are for the 776 lock. The atomic resonance of interest is monitored using the monitor vapour cell, interference filter (IF) and PD5. The rubidium energy levels used are indicated on the top right.}
	\label{setup}
\end{figure}

\section{Introduction}
Stabilising the optical output frequency of a laser, commonly known as laser locking, is essential for many areas of research. This is particularly true in atomic physics where the required absolute stability, being dictated by the width of atomic resonance lines, can often be sub-MHz. 
A plethora of methods have been developed for on- or near- resonant locking. Recent interest in performing thermal vapour experiments in the hyperfine Paschen-Back (HPB) regime \cite{Olsen2011, Sargsyan2014, Sargsyan2015, Sargsyan2015c, Sargsyan2018, Whiting2015,Whiting2016,Whiting2017a,Whiting2017b}, where the atomic resonances are typically Zeeman-shifted by tens of GHz, necessitates new methods of laser locking. In this Letter, we demonstrate a novel method for laser locking to a Zeeman-shifted two-photon transition.

The currently available methods for on- or near- resonant locking include locking to stable optical cavities \cite{Drever1983}, wavelength meters \cite{Kobtsev2007} 
and beat-note locks \cite{Day1992, Uehara2014}. 
In atomic physics research, lasers are often stabilised to a particular atomic resonance line.  
A variety of spectroscopic techniques can be used to generate a dispersive error signal with a zero-crossing at the lock-point.
These include frequency-modulation (FM) \cite{Bjorklund1980} and modulation transfer (MT) spectroscopy \cite{Shirley1982,McCarron2008}, which require external modulation of the laser to generate the dispersive lock signals. Other methods such as polarization spectroscopy \cite{Pearman2002}, saturated absorption spectroscopy \cite{hanes1969}, dichroic atomic vapor laser locking (DAVLL) \cite{Corwin1998, Petelski2003,Harris2008, Becerra2009} and prismatic deflection \cite{Purves2004}  do not require external modulation and are therefore experimentally simpler to implement. 
The Faraday effect can also be exploited to form an off-resonance laser lock \cite{Zentile2014a}, as can Zeeman-shift based locking (ZSAR) \cite{Reed2018}, both of which have the advantage of being tunable over a relatively wide range. 
\begin{figure*}[t!]
	\centering
	
	\includegraphics[width=2\columnwidth]{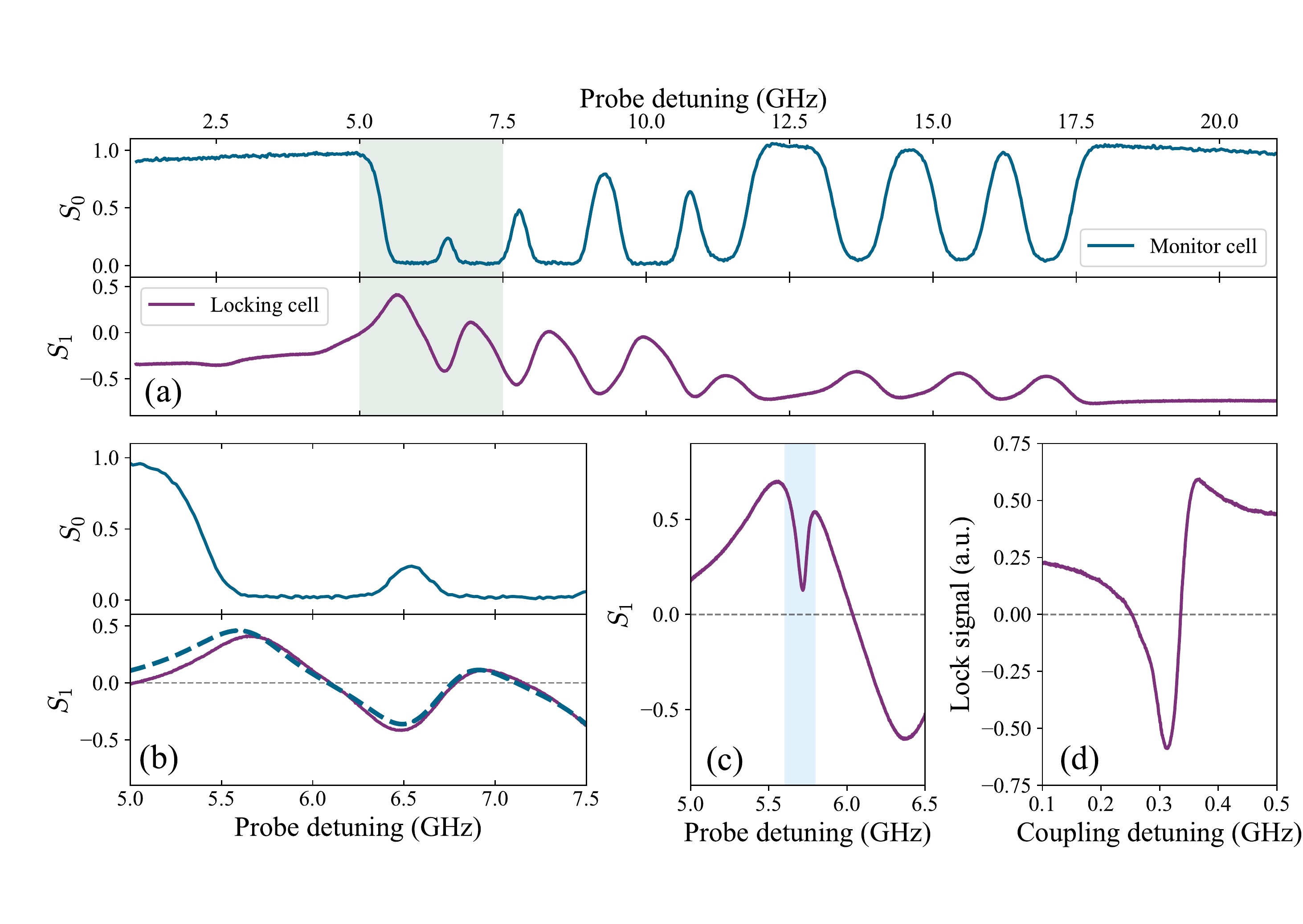}
	\caption{		
		\textbf{(a)} The monitor cell spectrum (top) is used to locate the required zero crossing on the locking cell signal (bottom). With only probe beam 1 on, the top panel shows the 780~nm spectrum in the hyperfine Paschen-Back regime in the monitor cell at 106$^{\circ}$C and bottom panel shows the S$_1$ signal in the locking cell at 100$^{\circ}$C. This temperature is a compromise between increased signal and optical depth and line broadening. Zero probe detuning is the weighted D$_2$ line centre of naturally abundant rubidium in zero magnetic field \cite{Siddons2008}.
		\textbf{(b)} The $S_1$ signal from probe beam 1 has a zero crossing that can be used in (d) to lock the probe laser. Shown is the shaded region from (a) where the dashed line is an ElecSus fit \cite{Zentile2015, Keaveney2017, Hughes2010} .  
		\textbf{(c)} The $S_1$ signal from probe beam 2 has an EIT feature (shaded region) when the coupling beam is turned on and resonant.  
		\textbf{(d)} With the probe laser now locked (using the method shown in part (b)), scanning the coupling laser gives the error signal used to lock the coupling laser.
	}
	\label{fig:spectra}
\end{figure*}
The laser frequency can also be stabilised at large detunings using saturation absorption spectroscopy \cite{Sargsyan2014Saturated} and a low-quality cavity technique \cite{Barboza2016}.

Alternatively, lasers can be intrinsically stabilised by placing atomic media in the external cavity feedback, obviating the need for external locking \cite{Keaveney2016}.
In multi-level atomic systems coupled with several lasers, the excited-state transitions can also be used as locking signals with some variations on the techniques above. These include locks based on electromagnetically induced transparency (EIT) \cite{Abel2009, Bell2007, Sargsyan2009Efficient},  fluorescence detection \cite{Kalatskiy2017}, and excited-state polarisation spectroscopy with \cite{parniak2016magneto} and without \cite{Carr2012} a small magnetic field. As our experiments are done in the HPB regime, none of the aforementioned techniques are immediately suitable for our purposes. Furthermore, quantum optics experiments \cite{Noh2011, Noh2015, Lee2016c, Lee2017, Srivathsan2013, Huber2014} often require long-integration times \cite{Willis2011a}, 
meaning that lasers may need to remain locked for several hours.

Here we present a technique to lock simultaneously two lasers to two transitions that form a ladder-type excitation scheme (see Fig. \ref{setup}), thus stabilising the sum of their frequencies over a timescale of hours, in a thermal vapour of $^{87}$Rb in the presence of a large magnetic field (0.6 T). This field is both large enough to gain access to the HPB regime and for the Zeeman shift to exceed the Doppler width. Both transitions are significantly Zeeman-shifted from their zero-field frequencies. Beyond being able to work in large fields, other advantages of our scheme include tunability on the first step of the excitation and the lock compensating for drift in one laser by automatic adjustment of the other.

\section{Concept}
We first use the off-resonant Faraday-rotation method described in \cite{Zentile2014a} to stabilise the 780~nm probe laser and then use a novel Faraday EIT method to stabilise the 776 nm coupling laser. 
In order to create a suitable crossing for the resonances of interest to us, our simultaneous two-photon resonant 
optical laser lock (STROLL) is implemented in the HPB regime.

Measurement of the optical rotation due to the Faraday effect near an atomic resonance provides an error-signal to which the probe laser can be locked \cite{Zentile2014a}. For an atomic medium in an external axial magnetic field, there are different refractive indices for right- and left-handed circularly polarised light (circular birefringence). This leads to the rotation of the plane of polarisation of linearly-polarised input light, where the rotation angle is proportional to the real part of the difference in refractive indices. The degree of rotation depends on the detuning of the light from the atomic resonances allowing the laser to be stabilised. The rotation is measured using the Stokes parameter, {$S_1 = (I_x - I_y)/I_0$}, which is the normalised difference in the intensities of orthogonal linear polarisation components of the output light (see Fig. \ref{setup}). The denominator $I_0$, the incident intensity,  normalises the definition so that $S_1$ lies between -1 and 1 
and so the lineshape of $S_1$ has a zero crossing. As shown on the bottom panel of \mbox{Fig. \ref{fig:spectra}(b)}, we can use an appropriate zero crossing in $S_1$ as an error signal to the feedback loop of our 
probe laser PID (proportional-integral-derivative) controller.

With the 780~nm probe laser locked, we can now lock the 776~nm coupling beam. 
We use a second probe beam that is overlapped with the coupling beam in the locking cell whilst ensuring that the first probe beam is not overlapped with the coupling beam. This configuration of beams leads to the presence of an EIT feature. EIT is a well-used technique in multi-level atomic systems \cite{Fleischhauer2005, Moon2011a}. It describes the reduction in the absorption of a weak probe laser when a strong coupling laser field is used to drive a resonant transition in a three-level atomic system, where the two resonant transitions are coherently coupled to a common state. 
Associated with the change in absorption, EIT results in a concomitant modification of the refractive index \cite{Budker1999}. In the HPB regime, the EIT feature only couples to one transition and so only changes the refractive index of one hand of polarisation -- hence EIT causes additional birefringence and a change in the $S_1$ signal \cite{Zentile2014a}.
EIT appears as a dispersive feature on the $S_1$ signal when the probe laser is scanning and the coupling beam is on and at a fixed frequency (See Fig. \ref{fig:spectra}(c) highlighted region). When the probe laser is locked and the coupling beam is scanning, we use this feature as the error signal (See Fig. \ref{fig:spectra}(d)) to the PID feedback loop of our 
coupling laser controller.   

\begin{figure}[t]
	\centering
	\mbox{\includegraphics[width=\linewidth]{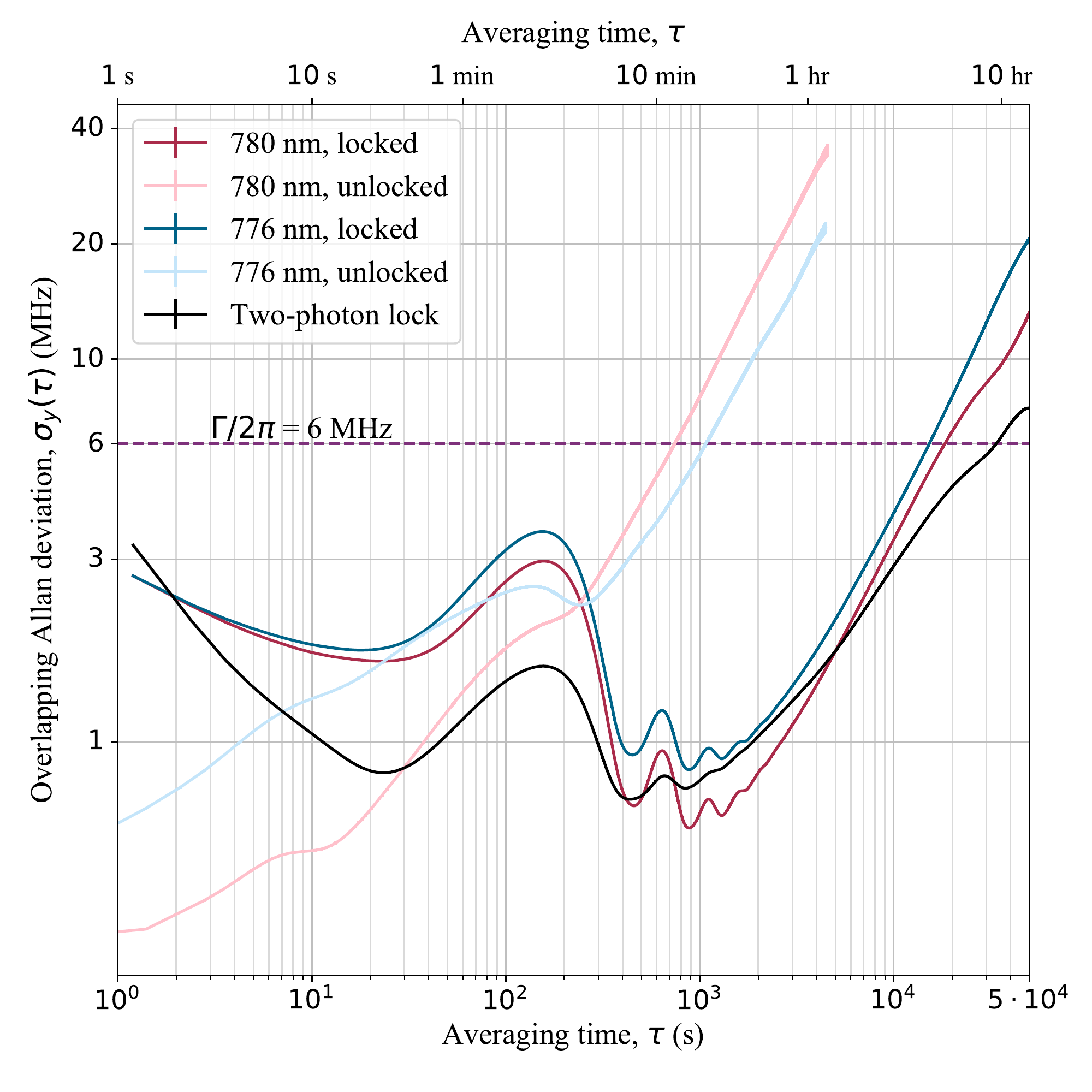}}
	\caption{Overlapping Allan deviation of the frequency measurement of the 780 nm probe laser, the 776 nm coupling laser and the summed frequency. $\Gamma$ is the natural linewidth of the \mbox{5S $\rightarrow$ 5P} probe transition.}
	\label{allan-deviation.PNG}
\end{figure}

\section{Experimental demonstration}
Fig. \ref{setup} shows a schematic of the experimental setup. Two weak (50 \textmu W) probe beams are focussed to a beam ellipse with waists of  \mbox{83$\pm$2 \textmu m $\times$ 106$\pm$2 \textmu m} (measured using \cite{Keaveney2018}) through a 1 mm-long heated vapour cell (the \lq locking  vapour cell\rq) of isotopically enriched rubidium (>98 \% $^{87}$Rb). A strong (16 mW) 776 nm coupling beam is focussed to a beam ellipse with waists of \mbox{74$\pm$2 \textmu m $\times$ 80$\pm$2 \textmu m} \cite{Keaveney2018} and counterpropagated through the cell. The second probe beam 
and the coupling 
beam, which are resonant with the 
\mbox{$|$5S$_{1/2}$, $m_J = \frac{1}{2}\rangle \rightarrow |$5P$_{3/2}$, $m_J = \frac{3}{2}\rangle$} and 
\mbox{$|$5P$_{3/2}$, $m_J=\frac{3}{2}\rangle \rightarrow |$5D$_{5/2}$, $m_J = \frac{1}{2}\rangle$} transitions respectively, are overlapped within the same cell.  We note that an advantage of our scheme is that most of the strong coupling beam can be reused in further experiments. The raw intensity differences (i.e. {$I_x-I_y$}) are generated with the use of polarising beam splitters (PBS) and photodiodes (PD1 \& PD2 for the probe lock and PD3 \& PD4 for the coupling lock).

The probe and coupling light are also sent through a 2 mm-long heated vapour cell (the `monitor vapour cell') of isotopically enriched rubidium (>98 \% $^{87}$Rb). Monitoring the absorption in this cell using a photodiode (PD5) allows choice over where to lock the 780 nm laser. This can be seen on Fig. \ref{fig:spectra}(a) where the zero crossing on the bottom panel is chosen depending on which resonance from the top panel is of interest. Both cells contain unknown buffer gas which causes an additional broadening of 7 MHz on the D$_2$ line.

Across each vapour cell, two cylindrical NdFeB magnets---Fig. \ref{setup} shows a cross-sectional view of the top-hat-profile of the magnets---are used to achieve a magnetic field of up to 0.6 T. The strength of each field can be varied by changing the separation of the respective magnets. By changing the strength of the field across the locking cell, we have tunability for the lock-point of the 780~nm laser, although the STROLL will remain locked to the two-photon resonance. The field over the region occupied by the 2 mm vapour cell has an rms variation of 4 \textmu{}T. Further details of field uniformity and magnet design can be found in \cite{whitingthesis2017} and \cite{zentile2015applications}.

To monitor the long-term stability of the locked lasers, Fig. \ref{allan-deviation.PNG} shows the overlapping Allan deviation \cite{NIST2008, Allan1966} of the concurrent frequency measurement of the 776 nm and 780 nm diode lasers where we have used a High Finesse WS7 wavemeter with a switcher box to simultaneously monitor both laser frequencies over a period of 24 hours. The lasers remain locked for the whole of this period.
The frequency instability of the sum of both lasers when locked is less than than the natural linewidth of the probe transition of 6 MHz and of the EIT linewidth of ~25 MHz. It is clear that, for most timescales, the frequency instability of the sum is less than the frequency instability of either the 780 nm laser or the 776 nm laser alone. STROLL keeps the lasers locked to the two photon transition: although the frequency of one laser may drift, the frequency of other changes accordingly to compensate. When unlocked, the lasers stay at an equivalent stability only for averaging times less than ~15 mins. This becomes important for quantum optics measurements where data must be accumulated over hours \cite{Whiting2017a} e.g. g$^{(2)}$ autocorrelation measurements.

Further improvements to the stability could be achieved if desired by adding active temperature stabilisation. In this work, the cell temperature was measured to be stable to better than 1$^{\circ}$C over several hours; this measurement also revealed that the peaks at 150 s in Fig. \ref{allan-deviation.PNG} are almost certainly due to temperature variation. The laser frequency stability achieved is sufficient for our purposes. However, temperature sensitivity is a known issue with Faraday locking \cite{Zentile2014a} with a temperature dependence of the zero crossing of 
$<$ 1 MHz/$^{\circ}$C.

The tunability of the probe laser lock-point is set by the strength of the magnetic field which gives several GHz of freedom. Further freedom arises from the presence of many possible zero crossings in the $S_1$ signal (because the Zeeman shift exceeds the Doppler width) as is seen in Fig. \ref{fig:spectra} (a) and from our PID electronics allowing us to choose different setpoint voltages.

\section{Conclusion}
We have demonstrated a technique to lock simultaneously two laser frequencies to the two photon transition, 5S$_{1/2}$ $\rightarrow$ 5D$_{5/2}$ in $^{87}$Rb in the presence of an applied magnetic field that is large enough to gain entry to the HPB regime. When locked simultaneously, we showed a frequency instability for the sum frequency of less than 6 MHz for nearly all measured time scales. Whilst in this paper the specific application was in rubidium, the concept is easily transferrable to three-level ladder systems in other alkali metals.

The authors acknowledge funding from EPSRC (Grant Nos. EP/L023024/1 and EP/R002061/1). F.P.-O. acknowledges the support of a Durham Doctoral Scholarship. The data-sets generated during and/or analysed during the current study are available in the Durham University Collections repository (doi:10.15128/r1ng451h522).


\bibliography{STROLL}

\bibliographyfullrefs{STROLL}

\end{document}